\documentstyle[prl,twocolumn,epsf,floats,aps]{revtex}

\begin{document}
\draft

\twocolumn[\hsize\textwidth\columnwidth\hsize\csname
@twocolumnfalse\endcsname

\title{Paired fractional quantum Hall states and the $\nu=5/2$ puzzle}
\author{N. Read}
\address{Department of Physics, Yale University, P.O. Box 208120, New Haven, CT
06520-8120}
\date{\today}
\maketitle

\begin{abstract}
Work on the problem of the $\nu=5/2$ fractional quantum Hall state
is reviewed, with emphasis on recent progress concerning paired
states.
\end{abstract}





\pacs{ } ]


\section{Introduction}

Recently, there has been a resurgence of interest in the quantized
Hall state at $\nu=5/2$ (and $7/2$) in single-layer samples. This
has been a long-standing puzzle, because it violates the
``odd-denominator rule'' satisfied by other observed quantized
Hall states. Moreover, the history of the problem exhibits some
remarkable twists and turns. In this paper, I review both early
and recent work on this problem, with emphasis on paired quantum
Hall states. There are now many definite theoretical predictions,
and a clear need for more experimental results. This paper is in
three main sections. Section II reviews the history of
experimental and theoretical work designed to reveal the nature of
the $\nu=5/2$ state, with the conclusion that it is theoretically
expected to be the Moore-Read state. Section III reviews my recent
work (with D. Green) which sheds light on the properties of this
state and other proposals. Section IV suggests experiments which
could pin down the nature of the observed state. Section V is the
conclusion.

\section{Quantized state at $5/2$}

\subsection{Experimental discovery}

In 1987, Willett and co-workers published results that showed
clear evidence of a quantized Hall plateau forming at low
temperatures at a filling factor of $5/2$ \cite{willett}.
Subsequently, it was shown \cite{eisenstein} that when a parallel
component $B_\parallel$ of the magnetic field is applied, the dip
in $\rho_{xx}$ decreases and disappears at a critical value.
Beyond this point, there is no quantized Hall state. This was
supported by activation energy gaps from $\rho_{xx}$ \cite{jim}.
The exactness of the quantization of $\rho_{xy}$ at low
temperature, and the smallness of the activation gaps, has been
confirmed in later work, in particular Ref.\ \cite{pan2}. The 5/2
state was the first even-denominator quantized Hall state
observed, and $5/2$ and $7/2$ remain the only ones observed in
single-layer samples.

In the early days, it was widely believed that reversed spins were
involved in the $\nu=5/2$ quantized Hall state. To understand this
point, we note that we will assume throughout that the lowest
(Landau level index $n=0$) Landau level (LL) is filled with
electrons of both spins. The remainder of the filling factor
$\nu=2+1/2$ is made up of electrons in the first excited ($n=1$)
LL. These electrons in the topmost, partially-filled or
``valence'', LL may be either fully-polarized and aligned with the
Zeeman effect of the magnetic field, or unpolarized, half of them
with spin up, half with spin down, or somewhere in between. We
will refer to these possibilities as ``polarized'',
``unpolarized'', and ``partially-polarized'', respectively, even
though the lowest LL is unpolarized in all cases. (An identical
discussion applies to the $7/2$ case, in which the lowest LL is
filled with both spins and the $n=1$ LL is half-filled with {\em
holes}. This is expected to exhibit similar physics, due to
particle-hole symmetry within a LL. Since this expectation appears
to hold up experimentally and theoretically, we neglect $7/2$
hereafter.) The main reasons for believing that the ground state
was either unpolarized or partially-polarized were (i) the 5/2
state occurred at relatively small magnetic fields (around 5 T)
where the Zeeman term is relatively small compared with Coulomb
interactions, and (ii) the parallel magnetic field increases the
Zeeman term, and in partially- or unpolarized states, this could
eventually cause a transition to a polarized ground state, which
might be an unquantized state, and such a transition was observed
\cite{eisenstein,jim}. Also, (iii) many types of
not-fully-polarized ground states would have charged excitations
with reversed spins relative to the ground state, such that the
energy gap would decrease in a characteristic way involving the
$g$ factor with $B_\parallel$; this agreed approximately with
observation \cite{jim}. However, none of these arguments is
irrefutable, and we will see in a moment that there are
alternative possible explanations of the observations.

The quantized state at $5/2$ may be contrasted with the vicinity
of half-filling of other LLs. At $\nu=1/2$ and $3/2$, an
unquantized, compressible, Fermi-liquid-like state is observed
\cite{hlr}. On the other hand, at half-filling of higher LLs,
$\nu=9/2$, $11/2$, \ldots, a highly anisotropic, unquantized,
compressible state has been observed \cite{lilly,du}. Such
``stripe'' states for half-filling LLs of high index were
predicted theoretically in advance \cite{koulakov,moessner}. The
effect of $B_\parallel$ on $\nu=5/2$ was then reexamined, and it
was found that the compressible state at $B_\parallel$ greater
than the critical value is again an anisotropic or ``stripe''
state \cite{pan,lilly2}.

While doubts about the unpolarized nature of the 5/2 state had
begun to surface throughout the 90's, the last results
\cite{pan,lilly2} gave a very strong indication that the physics
of the 5/2 state had not been understood. The parallel field
experiments did not directly probe the polarization of the state,
yet the transition to a stripe state (and the sensitivity of the
spatial orientation of the anisotropy, for various $\nu$ where it
is observed, to the direction of $B_\parallel$) indicates that the
interactions between the electrons are being affected by
$B_\parallel$, which is possible because of the finite thickness
of the electron wavefunctions in the two-dimensional electron gas.
This mechanism is then a plausible alternative to the explanation
via the increased Zeeman term, so that the 5/2 state at
$B_\parallel=0$ might be fully polarized after all.

\subsection{Theory for incompressible states at even denominators}

The ``odd-denominator'' rule, that quantized Hall states are
observed only at filling factors with odd denominators, dates from
the earliest observations of the fractional quantum Hall effect,
and was ``explained'' by Laughlin's theory \cite{laugh} for
$\nu=1/q$ ($q$ odd) and its extension to other filling factors by
the hierarchy and composite-fermion theories \cite{hier,jain}
(which are essentially equivalent \cite{read90,blokwen}); these
approaches produce all, and only, odd-denominator fractions, and
cannot account for the 5/2 state.

Some early attempts to generalize Laughlin's results in different
directions were put forward by Halperin \cite{halp83}. One idea
was that if, for some reason, electrons are bound in pairs, then
these pairs are charge 2 bosons (throughout, we give charges in
units of that on the electron), and these can in principle form a
Laughlin state for bosons. The filling factor for the electrons in
such a state is then of the form $\nu=4/m$, where $m$ is even;
this gives a sequence of fractions that includes $1/2$, $1/4$,
\ldots. Such a state has Laughlin quasiparticle excitations of
charge $\pm 2/m=\pm \nu/2$. Such excitations are a common feature
of all the paired states we will discuss; note that the
quasiparticle charge is fractionalized compared with the usual
Laughlin states, which give charge $\pm\nu$ quasiparticles.
Excitations obtained by breaking the electron pairs are here
assumed to be very costly in energy. In applying these and the
following states to the 5/2 problem, we again assume that the
$n=0$ Landau level is filled with both spins, and use the fact
that the $n=1$ Landau level can be mapped to the $n=0$ Landau
level, so we will describe it as $\nu=1/2$. Note, however, that
the interaction Hamiltonian should be that for electrons in the
$n=1$ LL (and possibly should include the effect of virtual
excitations involving the $n=0$ LL).

After the experiment of Willett {\it et al.} \cite{willett},
Haldane and Rezayi (HR) \cite{hr} investigated spin-singlet (i.e.\
unpolarized) states at $\nu=1/2$. They used the pairing idea, but
for spin-singlet pairs, which allows two electrons to occupy the
same single particle state in a LL. They also used a ``hollow
core'' Hamiltonian, in which the zeroth Haldane pseudopotential
$V_0$ (which corresponds to the contact interaction of two
electrons) is zero, motivated by the reduction of this parameter
in the $n=1$ LL. For a hollow core model in which only the first
pseudopotential $V_1>0$ is nonzero, they found a unique exact
ground state at $\nu=1/2$, which was argued to be incompressible
(similar Hamiltonians and ground states exist for $\nu=1/4$,
$1/6$, \ldots). The nature of the HR state will be discussed again
later.

Subsequently, Moore and Read (MR) \cite{mr} pointed out that
paired states at $\nu=1/q=1/2$, $1/4$, \ldots, can be interpreted
as BCS pairing \cite{bcs} of composite fermions in zero net
magnetic field. In this point of view, the HR state is a
spin-singlet d-wave (d$_{x^2-y^2}-i$d$_{xy}$) pairing state of
composite fermions; the spin-singlet nature of the state, which
was initially obscure \cite{hr}, becomes obvious from this point
of view \cite{mr}. Inspired by this picture and by the structure
of the HR state, MR constructed another state, a p-wave
(p$_x-i$p$_y$) polarized state, which they called the Pfaffian
state. Motivated by deep considerations of conformal field theory
and its relation to the quantum Hall effect, they argued that the
charge $1/2q$ excitations of the MR state obey nonabelian, rather
than ordinary fractional, statistics. We will explain the meaning
of this later.

Soon after, Greiter and coworkers \cite{gww} considered the MR
state, also from the viewpoint of pairing composite fermions, but
argued that the statistics are ordinary abelian fractional
statistics, based on the Halperin picture of a Laughlin state of
charge 2 bosons (the resolution of this issue will be discussed
below). They found a three-body Hamiltonian for which the MR state
is the unique exact zero-energy eigenstate for the case of charge
1 bosons (instead of electrons) at $\nu=1$ (the generalization to
electrons at $\nu=1/2$ in Ref.\ \cite{gww} is incorrect, but was
corrected by later authors; for still smaller $\nu=1/q$, two-body
terms are necessary in addition \cite{rr2}); this Hamiltonian was
very useful in later work. Finally, they suggested that the MR
state may represent the observed $\nu=5/2$ state, which should
therefore be polarized, in conflict with the conventional wisdom
at the time.

Numerical work on the $\nu=5/2$ problem absorbed many person-years
of labor during the 1990's. Eventually, results were forthcoming
that now strongly indicate that the 5/2 state is expected to be
the spin-polarized MR state. The first published work was by Morf
\cite{morf}, who argued that the ground state is spin-polarized,
with a large overlap with the MR state in finite size systems.
Further, weakening the short range repulsion led to a transition
to a compressible state, while strengthening it gave a transition
to a Fermi-sea state, like that in the lowest Landau level
\cite{hlr,rr1}. Rezayi and Haldane \cite{rh} confirmed Morf's
results, using the torus geometry, rather than the sphere, and
studied the transitions in detail. In particular, they established
that the compressible state at weak short-range interaction is a
stripe state \cite{koulakov,moessner,rhy}. Also, results for
realistic potentials, including finite thickness effects,
screening by the $n=0$ LL, and tilted magnetic field, confirm that
at $B_\parallel=0$, the ground state at $5/2$ should be the MR
state, while $B_\parallel$ drives the system into a stripe state
\cite{rh}, as occurs experimentally \cite{pan,lilly2}. Apparently,
the system with $B_\parallel=0$ lies close enough to the
transition to the stripe state that a small change in the
interactions due to nonzero $B_\parallel$ (through finite
thickness effects) can push it into the stripe phase. From this
work, a systematic picture has emerged of how at half-filling of
each LL with $B_\parallel=0$, the ground state evolves
successively from Fermi-liquid (for $n=0$), to paired (for $n=1$),
to striped (for $n>1$).

There are also Monte Carlo studies that indicate a low energy for
the MR state at $\nu=5/2$ \cite{pmbj}. A recent attempt to show
that Cooper pairs of composite fermions form using trial states
\cite{spj} has been criticized \cite{read00}.

\section{BCS pairing of composite fermions}

\subsection{MR state}

Now that we have seen strong reasons to believe that the $\nu=5/2$
state is the MR state, we are motivated to inquire more deeply
into its properties, in search of experimental signatures. In this
section, we review recent progress in understanding these
properties \cite{rg}.

First, we will simply say that for $\nu=1/2$, a composite fermion
is an electron bound to two vortices in the wavefunction of the
other electrons (see e.g.\ \cite{read94}). This object is a
fermion, is electrically neutral, and experiences zero effective
magnetic field $B_{\rm eff}$---{\em each} of these properties
holding {\em only} at $\nu=1/2$. These statements generalize to
$\nu=1/q$ for electrons bound to $q$ vortices, $q$ even; if
instead $q$ is odd, the statements hold, except for the
statistics: the object is a composite boson
\cite{g,gm,read87,zhk}.

Fermions in zero magnetic field can form a ground state
represented by a BCS trial wavefunction (formally, we arrive at
this by a mean field approximation that yields fermions in zero
net magnetic field \cite{hlr}, followed by the BCS mean field
\cite{bcs} approximation that describes pairing). In the quantum
Hall context, such a state is an incompressible fluid that
generalizes the Laughlin state, in the following sense (related to
that of Ref.\ \cite{halp83}). The Laughlin state can be viewed as
a Bose condensate of composite bosons \cite{g,gm,read87,zhk}. The
condensate allows magnetic flux (more accurately, vortices) to be
inserted only in quantized amounts that cost a nonzero, finite
energy (the Meissner effect for the condensate); the quantum Hall
relation between flux and charge implies that these excitations
carry a charge $\pm 1/q$, and since they cost finite energy, the
state is incompressible. In the case of the paired states at
$\nu=1/q$, the Cooper pair condensate carries twice the electric
charge, which halves the flux quantum; the fluid is again
incompressible, but the elementary charged excitations carry
charge $\pm 1/2q$ \cite{mr}.

The wavefunction written down by MR for one possible
spin-polarized state with filling factor $1/q$ ($q$ even) was
\begin{equation}
\Psi_{\rm MR}(z_1,\ldots,z_N)={\rm
Pf}\,\left(\frac{1}{z_i-z_j}\right)\prod_{i<j}(z_i-z_j)^q,
\end{equation}
where we have used complex coordinates $z_j=x_j+iy_j$ for the $N$
electrons ($N$ even), omitted the ubiquitous Gaussian factor, and
the Pfaffian $\rm Pf$ is defined by
\begin{equation}
{\rm Pf}\,\left(M_{ij}\right)={\cal A}\left( M_{12}M_{34}\ldots
M_{N-1,N}\right),
\end{equation}
where $M_{ij}$ are the elements of an antisymmetric matrix, and
$\cal A$ denotes the operation of antisymmetrization, normalized
such that each distinct term appears once with coefficient 1. As
we will see, the Pfaffian is the general structure of the
position-space form of the BCS state in the spin-polarized, p-wave
case, so the wavefunction $\Psi_{\rm MR}$ represents BCS pairing
of composite fermions \cite{mr}.

Pairing composite fermions leads us to expect two types of
elementary excitations of this ground state. One type are the
charged vortices discussed above, with charge $1/2q$, which
according to MR are supposed to obey nonabelian statistics. The
other type are the analog of the BCS quasiparticles, which are
fermions, and are created (in twos) by breaking pairs; there
should be an energy gap for these. These excitations are charge
zero, like the underlying composite fermions (since $\nu=1/q$).

To make further progress, we consider (following Ref.\ \cite{rg})
BCS theory at the mean field level for p-wave pairing of spinless
or spin-polarized fermions \cite{schrieffer}. We are not
interested here in the mechanism for an attractive interaction
between composite fermions that gives the pairing, nor in solving
the gap equation. Rather we are interested in the physical
properties of the resulting ground states, especially those
related to the fermionic quasiparticles. At the mean field level,
one works with the following effective Hamiltonian for the
quasiparticles:
\begin{equation}
K_{\rm eff}=\sum_{{\bf k}}\left[\xi_{\bf k} c_{\bf k}^\dagger
c_{\bf k} +\frac{1}{2}\left( \Delta_{\bf k}^\ast c_{-{\bf
k}}c_{\bf k}+\Delta_{\bf k} c_{\bf k}^\dagger c_{-{\bf
k}}^\dagger\right)\right],
\end{equation}where $\xi_{\bf k}=\varepsilon_{\bf k}-\mu$ and
$\varepsilon_{\bf k}$ is the single-particle kinetic energy and
$\Delta_{\bf k}$ is the gap function. For the usual fermion
problems, $\mu$ is the chemical potential, but may not have this
meaning in the quantum Hall applications. At small ${\bf k}$, we
assume $\varepsilon_{\bf k}\simeq k^2/2m^\ast$ where $m^\ast$ is
an effective mass, and so $-\mu$ is simply the small ${\bf k}$
limit of $\xi_{\bf k}$. For complex p-wave pairing, we take
$\Delta_{\bf k}$ to be an eigenfunction of rotations in ${\bf k}$
of eigenvalue (two-dimensional angular momentum) $l=-1$, and thus
at small ${\bf k}$ it generically takes the form
\begin{equation}
\Delta_{\bf k}\simeq\hat{\Delta}(k_x-ik_y), \end{equation}where
$\hat{\Delta}$ is a constant. For large ${\bf k}$, $\Delta_{\bf
k}$ will go to zero. The $c_{\bf k}$ obey $\{c_{\bf k},c_{{\bf
k}'}^\dagger\}=\delta_{{\bf k}{\bf k}'}$; for the moment we work
in a square box of side $L$.

The diagonalization of this Hamiltonian is a standard exercise.
The quasiparticle dispersion relation is
\begin{eqnarray}
E_{\bf k} &=& \sqrt{\xi_{\bf k}^2+|\Delta_{\bf k}|^2}\nonumber\\
&\simeq&\sqrt{\mu^2 + |\hat{\Delta}|^2 k^2}
\end{eqnarray}
as $k$, $\mu\rightarrow 0$. The corresponding ground state
wavefunction in position space for $N$ particles is
\begin{equation}
\Psi({\bf r}_1,\ldots,{\bf r}_N)={\rm Pf}\,\left(g({\bf r}_i-{\bf
r}_j)\right),
\end{equation}
where $g({\bf r})$ is the inverse Fourier transform of  $(\xi_{\bf
k}-E_{\bf k})/\Delta_{\bf k}^\ast$.

A transition occurs at $\mu=0$, where $E_{\bf k}$ becomes gapless
at ${\bf k}={\bf 0}$. The large $\bf r$ behavior of $g$ is
different on the two sides of, and at, this transition. For
$\mu>0$, $g({\bf r})\sim 1/z$ as ${\bf r}\to\infty$; for $\mu=0$,
$\sim 1/(z|z|)$; and for $\mu<0$, $\sim e^{-{\rm const.}r}|z|/z$.
Because of the long-range behavior of $g$, we call $\mu>0$ the
weak-pairing phase, while we call $\mu<0$, where the pairs are
tightly bound, the strong-pairing phase. Intermediate behavior is
found at the transition, $\mu=0$. We see that the weak-pairing
regime has the same aymptotic behavior as the MR state has for all
$\bf r$; this $\mu>0$ case also corresponds to weak attractive
coupling. We will argue that this phase generically has the
properties associated with the nonabelian statistics of the MR
state. The strong-pairing regime corresponds to very strong
attractive coupling, which can produce $\mu<0$, and we will see
that the physics there is that of the simple Halperin picture of a
Laughlin state of charge 2 bosons.

Next we consider the fermion spectrum in non-translationally
invariant situations, specifically edges and vortices. We allow
either $\xi_{\bf k}$ or $\Delta_{\bf k}$ to depend on position.
Again, in principle the form of both of these should be found by
solving the mean-field equations self-consistently, but we do not
do this here; the results we are interested in are generic
throughout a phase, and do not change unless a transition is
crossed, so self-consistency should not matter. We work close to
the transition, where $\mu$ is small, and assume the position
dependence is in each case slowly-varying, so that the small $k$
behavior is sufficient. The problem of finding the fermion
spectrum is simply the solution of the Bogoliubov-de Gennes
equation \cite{degennes}, and in this limit it reduces to a Dirac
equation, with $|\hat{\Delta}|$ as the speed of light \cite{rg}.
It has reality properties that imply that the quasiparticles are
their own antiparticles, and Dirac fermions with this property are
known as Majorana fermions.

For the case of an edge, we let $\mu$ depend on $x$, but not $y$
(for an edge parallel to the $y$-axis); $\mu$ becomes large and
negative outside the edge. This corresponds to a large, positive
potential for electrons outside the edge, which confines them to
the interior. Thus outside the system, we would be in the
strong-pairing phase, but with the particle density going to zero
far outside. If the interior is in the weak-pairing phase,
$\mu>0$, then the edge is effectively a domain wall that separates
regions in either phase. It is now described by a Dirac equation
with a mass $\mu$ that changes sign. It can be shown that there is
a gapless low-energy spectrum of chiral Majorana fermion modes
that are bound to, and propagate in one direction along, the
domain wall \cite{rg}, whereas the remainder of the spectrum,
associated with bulk states, has an energy gap. On the other hand,
when the interior is in the strong-pairing phase, there is no such
domain wall at the edge, and no gapless chiral modes are present.
The chiral Majorana fermions on the edge in the weak-pairing phase
agree with the results obtained earlier for the three-body
Hamiltonian \cite{gww} for which the MR state is the exact ground
state \cite{wen3,milr}. In addition, the fractional quantum Hall
state has gapless chiral density excitations, the usual ``edge
states'', which are not obtained from the fermion analysis.

Vortices may be thought of as small circular edges, enclosing a
half quantum of magnetic flux. When the bulk is in the
weak-pairing phase, a similar calculation shows that there is a
Majorana zero-energy state associated with each vortex (vortices
should occur in even numbers if the boundary conditions at
infinity are the same as for no vortices). When the separation of
the vortices is finite, these energies are split by amounts that
go to zero exponentially fast as the separations go to infinity.
Neglecting these splittings, the many-particle states have a
degeneracy $2^n$ for $2n$ vortices, or $2^{n-1}$ if we restrict to
a fixed particle number (either even or odd). This number arises
because only $n$ creation-annihilation operator pairs can be
formed from the $2n$ real (Majorana) fermion operators.  This
asymptotic degeneracy of many-particle states agrees with that
obtained \cite{rr2,nayak} for any separation of vortices using the
three-body Hamiltonian. Because the number is $2^n$, not $2^{2n}$,
it cannot be viewed as a two-fold degeneracy of each vortex, but
instead is somehow nonlocal; the states are shared among the
vortices. It is these facts that give rise to nonabelian
statistics. In the strong-pairing phase, there are no such zero
modes of the BdG equation, and the multivortex states are
nondegenerate for fixed positions, even asymptotically.

To summarize, we have found that the weak-pairing phase has
properties previously associated with the MR state, while the
strong-pairing phase has no such properties. The strong-pairing
phase is then left with only the properties it inherits from the
condensation of pairs, and is entirely consistent with the physics
that follows from Halperin's picture of a Laughlin state of charge
2 bosons \cite{halp83}.

Nonabelian statistics of the vortices (or quasiparticles) of the
MR or weak-pairing phase may now be explained. When all
separations are large, a set of $2n$ quasiparticles has $2^n$
degenerate ground states. When two of them are exchanged
adiabatically, the effect can be described as a matrix operation
on the space of possible states (this is derived simply in Ref.\
\cite{ivanov}), rather than just multiplication by a phase as in
the more familiar abelian ``fractional'' statistics. Matrices for
different exchanges will not commute, hence the name nonabelian.
We note that the order of limits (infinite separation before the
limit of low-speed exchange) is important here. In practice, for
finite separations, the exchange must not be done too slowly,
compared with the exponentially-small splittings. It has been
suggested in some quarters that these exchanges could be used to
perform computations in a quantum computer.

We will also comment here very briefly on the effects of disorder
on the MR state. In the quantum Hall effect, potential disorder
can nucleate and localize vortices (quasiparticles); since the
latter have finite energy, they can be localized in essentially
uncorrelated positions, with some mean density depending on the
disorder strength and on the distance in magnetic field (filling
factor) from the center of the quantized Hall plateau. In the case
of the MR phase, each vortex carries a fermion zero mode, and as
we have noted, the degeneracies of many-particle states can be
split by tunneling of the fermions from one vortex to another. For
a finite density of vortices, there will then be a band of
localized low-energy fermion excitations, like an ``impurity
band''. It has been argued \cite{rl} that this has the effect of
destroying the properties of the MR phase and replacing it by a
disordered version of the strong-pairing or Halperin phase. In
particular, the chiral Majorana fermion edge modes will be
destroyed, by backscattering and localization via the nearby
vortex zero modes in the bulk. However, for a high-quality sample,
all of this may be occurring at extremely low energies and large
length scales; thus in finite systems at finite temperature, the
only effect may be that there is a bath of the quasidegenerate
fermion states on the localized vortices. There could be a lot of
interesting physics associated with this.

\subsection{HR state}

We will consider here briefly the fate of the HR state. The
original HR state had (spinor-valued) wavefunction
\begin{equation}
\Psi_{\rm HR}={\rm
Pf}\,\left(\frac{\uparrow_i\downarrow_j-\downarrow_i\uparrow_j}{(z_i-z_j)^2}
\right)\prod_{i<j}(z_i-z_j)^q,
\end{equation}
(where $\uparrow_i$ means the spin state $\uparrow$ for particle
$i$, and the product is the tensor product) which corresponds to
spin-singlet (as in the original BCS theory \cite{bcs}) complex
d-wave pairing, and the corresponding gap function would have the
form
\begin{equation}
\Delta_{\bf k}\simeq \hat{\Delta}(k_x-ik_y)^2
\end{equation}
for small $k$. We may perform a similar analysis of BCS mean field
theory in this case, and there are again weak- and strong-pairing
phases separated by transition. However, the behavior of the
pairing function $g$ at the transition is $g\sim 1/z^2$ (times the
spin singlet factor), that is, the same as in the HR state. This,
together with an analysis of the ground states on the torus,
suggested that the hollow-core Hamiltonian, and the HR
wavefunction, are sitting right at the weak--strong-pairing
transition point, and hence the fermionic quasiparticle spectrum
in the bulk should be gapless, $E_{\bf k}\sim k^2$. Since this
would be reached in practice by tuning a parameter, it cannot be
the generic behavior of a phase---even if the 5/2 state is spin
unpolarized. Then earlier results on the edge and quasiparticle
properties of the HR state \cite{milr,rr2} are moot. Instead,
there is a weak-pairing phase which has abelian statistics, and is
equivalent to states obtained in several earlier approaches
\cite{rg}, and also a strong-pairing phase.

\section{Experiments needed}

Several difficult experiments, which have been done successfully
for some states in the lowest LL, would pin down the nature of the
5/2 state if they could be done. These are (i) measurement of the
spin polarization by Knight shift, to see if the valence Landau
level is polarized; (ii) shot noise or antidot experiments, to
measure the fractional charge of excitations, which should be
$1/4$ in a paired state at $\nu=5/2$; (iii) tunneling into the
edge, to measure the exponent in the current-voltage relation
$I\sim V^\alpha$, which (neglecting the lowest-Landau-level
contribution with $\alpha=1$) should be $\alpha=3$ in the weak-,
but $8$ in the strong-pairing phase \cite{milr}. Together,
positive results for these experiments would show that the state
is a spin-polarized weak-pairing phase, which must almost
certainly be the MR phase. However, the smallness of the gaps in
the 5/2 state \cite{jim,pan2} make all of these extremely
difficult. Of course, other suggestions for ways to probe the
physics of the MR state would be welcome.

\section{Conclusion}

To conclude, there is now plenty of theoretical evidence that the
5/2 state is the MR phase. The latter has a great deal of
fascinating physics, including chiral fermion edge excitations,
nonabelian statistics due to fermion zero modes on vortices, and
resulting effects of disorder. Clearly, more experiments are
needed to finally solve the puzzles posed by $\nu=5/2$, the most
surprising of fractional quantum Hall states.


\acknowledgments

This work was supported by the NSF under grant no.\ DMR-98-18259.

\end{document}